\documentclass{iau} 
\usepackage{graphicx}

\newcommand\rhalf{r_{\rm half}}

\newcommand\rhohalf{\rho_{\rm half}}

\newcommand\rstar{r_*}

\newcommand\Rstar{R_*}

\def\Sersic{{S\'ersic}}

\title[Origin of {\Sersic} and Einasto profiles] 
{On the origin of  {\Sersic} profiles of galaxies and Einasto profiles of dark-matter halos}

\author[Carlo Nipoti]   
{Carlo Nipoti}

\affiliation{Department of Physics and Astronomy, Bologna University,\\ viale Berti-Pichat 6/2, I-40127 Bologna, Italy, email: {\tt carlo.nipoti@unibo.it} \\{\tiny ~}\\{\rm Submitted, 28 April 2016}}

\pubyear{2016}
\volume{321}  
\jname{Formation and evolution of galaxy outskirts} %
\editors{A.G. de Paz, J. Knapen \& J. Lee, eds.}
\begin{document}

\maketitle

\begin{abstract}
The surface-brightness profiles of galaxies $I(R)$ and the density
profiles of dark-matter halos $\rho(r)$ are well represented by the
same analytic function, named after either {\Sersic}, $I\propto
e^{-(R/\Rstar)^{1/m}}$, or Einasto, $\rho\propto
e^{-(r/\rstar)^\alpha}$, where $\Rstar$ and $\rstar$ are
characteristic radii.  Systems with high {\Sersic} index $m$ (or low
Einasto index $\alpha$) have steep central profiles and shallow outer
profiles, while systems with low $m$ (or high $\alpha$) have shallow
central profiles and steep profiles in the outskirts. We present the
results of idealized numerical experiments which suggest that the
origin of these profiles can be traced back to the initial density
fluctuation field: high-$\alpha$ (low-$m$) systems form in smooth
regions via few mergers, while low-$\alpha$ (high-$m$) systems form in
clumpy regions via several mergers.


\end{abstract}

\firstsection 

\section{Einasto and {\Sersic} profiles}

The surface-brightness profiles of galaxies are successfully described
by the \cite{Ser68} law $I(R)\propto e^{-(R/\Rstar)^{1/m}}$, where
$\Rstar$ is a characteristic projected radius. The same analytic
function, written in the form $\rho(r)\propto
e^{-(r/\rstar)^{\alpha}}$, where $\rstar$ is a characteristic
intrinsic radius, and known as \cite{Ein65} profile, represents well
the density distribution of dark-matter halos in cosmological $N$-body
simulations (\cite[Merritt et al. 2005]{Mer05}).  The Einasto (or
{\Sersic}) function has the property that a single index ($\alpha$ or
$m$) determines the distribution at both small and large radii. When
$\alpha$ is low (or $m$ is high) the central profile is steep and the
outer profile is shallow, while when $\alpha$ is high (or $m$ is low)
the central profile is shallow and the outer profile is steep.
\cite{Cen14} has proposed that profiles in which the central and outer
slopes are anti-correlated arise naturally in the standard
cosmological model with initial density fluctuations represented by a
Gaussian random field (GRF). When the fluctuation field is dominated
by long-wavelength modes the system forms via a coherent collapse,
leading to a profile shallow in the center and steep in the
outskirts. When short-wavelength modes dominate, a steep dense central
core is initially formed and late infall of substructures produces an
extended envelope. Indications in this direction come from
cosmological $N$-body simulations: halos with higher values of
$\alpha$ are assembled more rapidly than halos with lower values of
$\alpha$ (\cite[Ludlow et al. 2013]{Lud13}). Similarly, binary merging
simulations show that the {\Sersic} index increases while a galaxy is
growing via dissipationless mergers (\cite[Nipoti et al. 2003]{Nip03};
\cite[Hilz et al. 2013]{Hil13}).

\begin{figure}[t!]
\begin{center}
\centerline{ \includegraphics[width=0.5\textwidth]{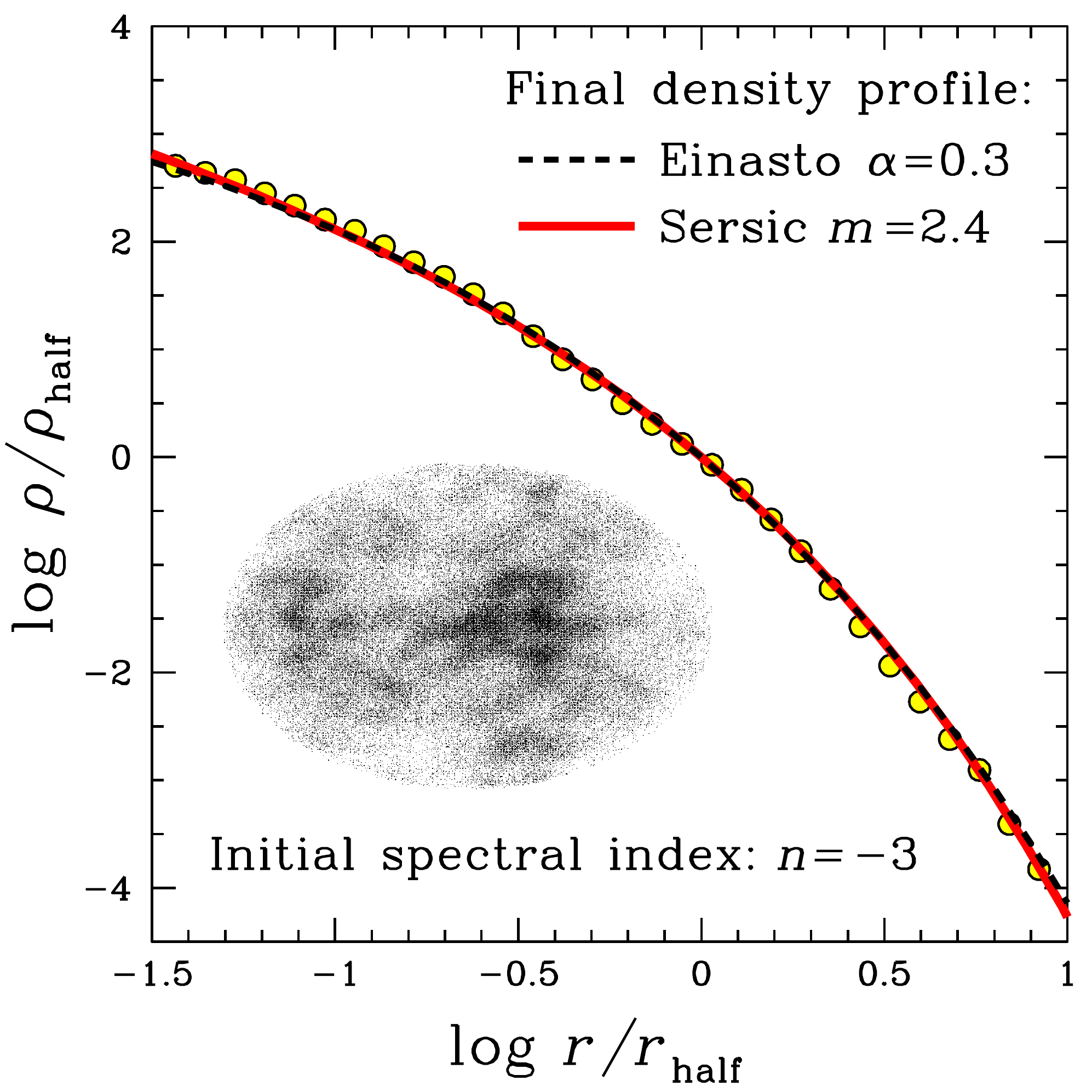} 
 \includegraphics[width=0.5\textwidth]{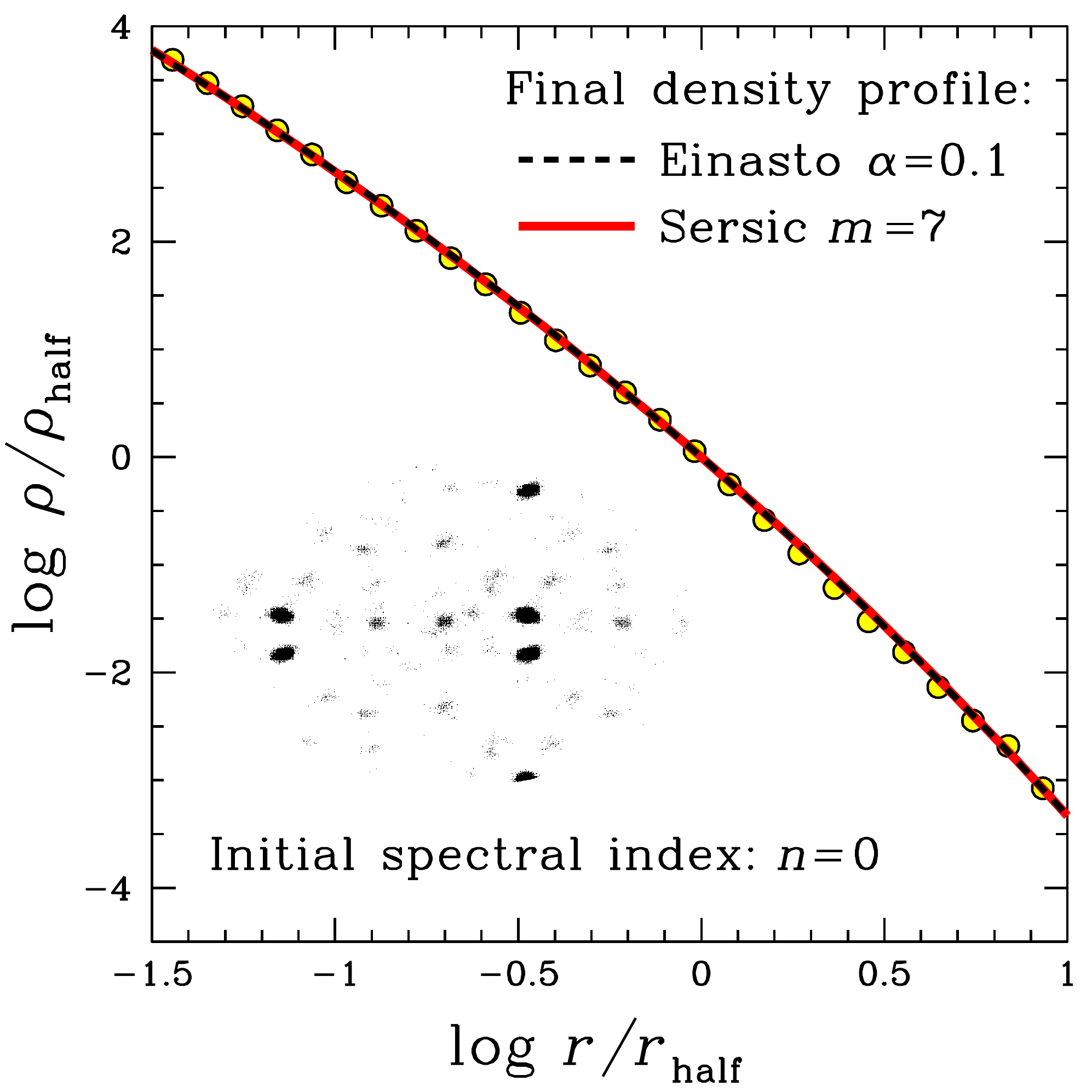} }
 \caption{Final angle-averaged density profiles (circles) with their
   best-fitting Einasto (dashed curves) and deprojected {\Sersic}
   (solid curves) profiles for two simulations with initial
   fluctuation power spectrum index $n=-3$ (left-hand panel) and $n=0$
   (right-hand panel; $\rhohalf$ is the density at the half-mass
   radius $\rhalf$).  These two simulations differ from the
   corresponding simulations presented in N15 only in the specific
   realization of the GRF.  In each panel the initial projected
   distribution of particles is shown as an inset.}
   \label{fig:den}
\end{center}
\end{figure}

\section{Numerical experiments}

\begin{figure}[t!]
\begin{center}
\centerline{ \includegraphics[width=0.5\textwidth]{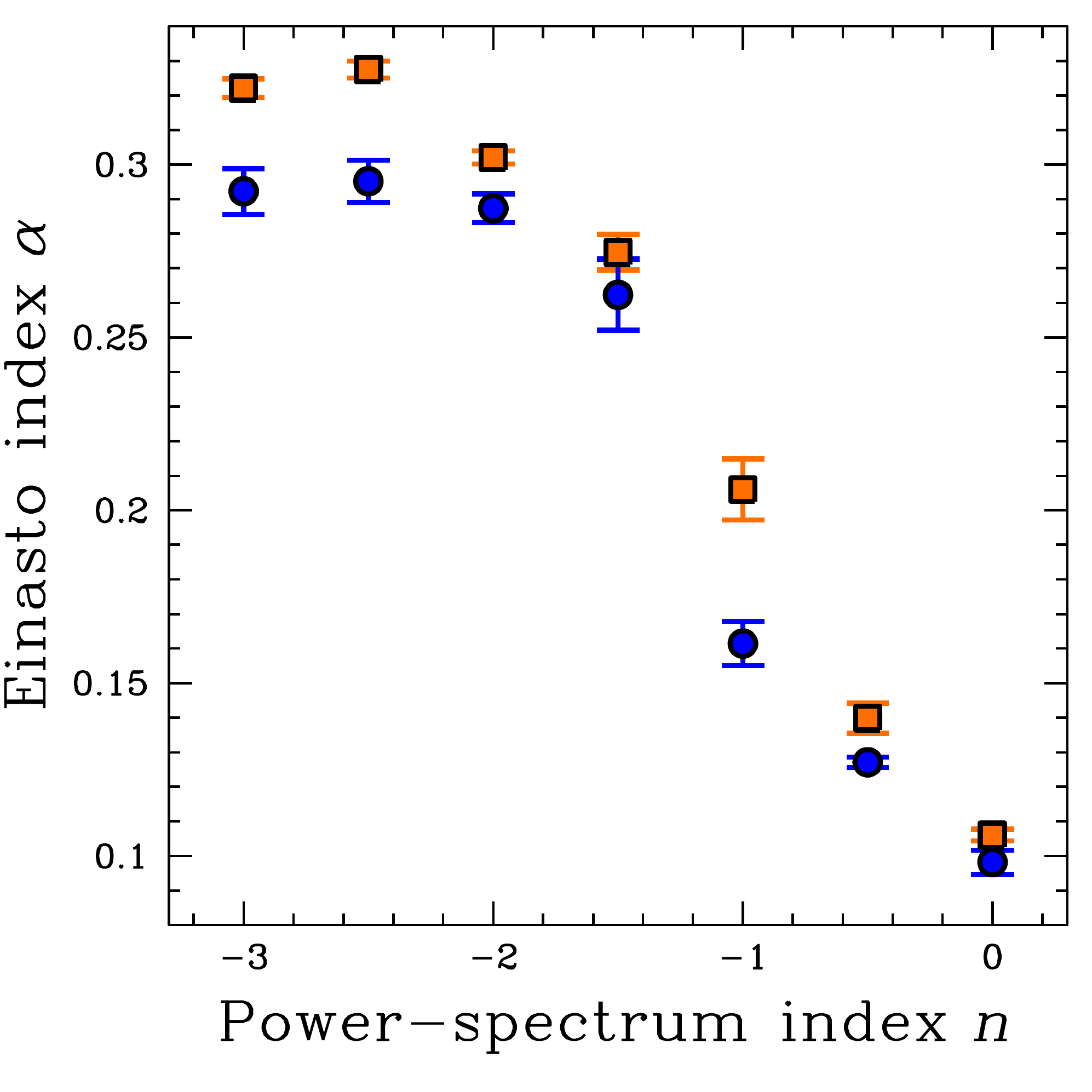}
  \includegraphics[width=0.5\textwidth]{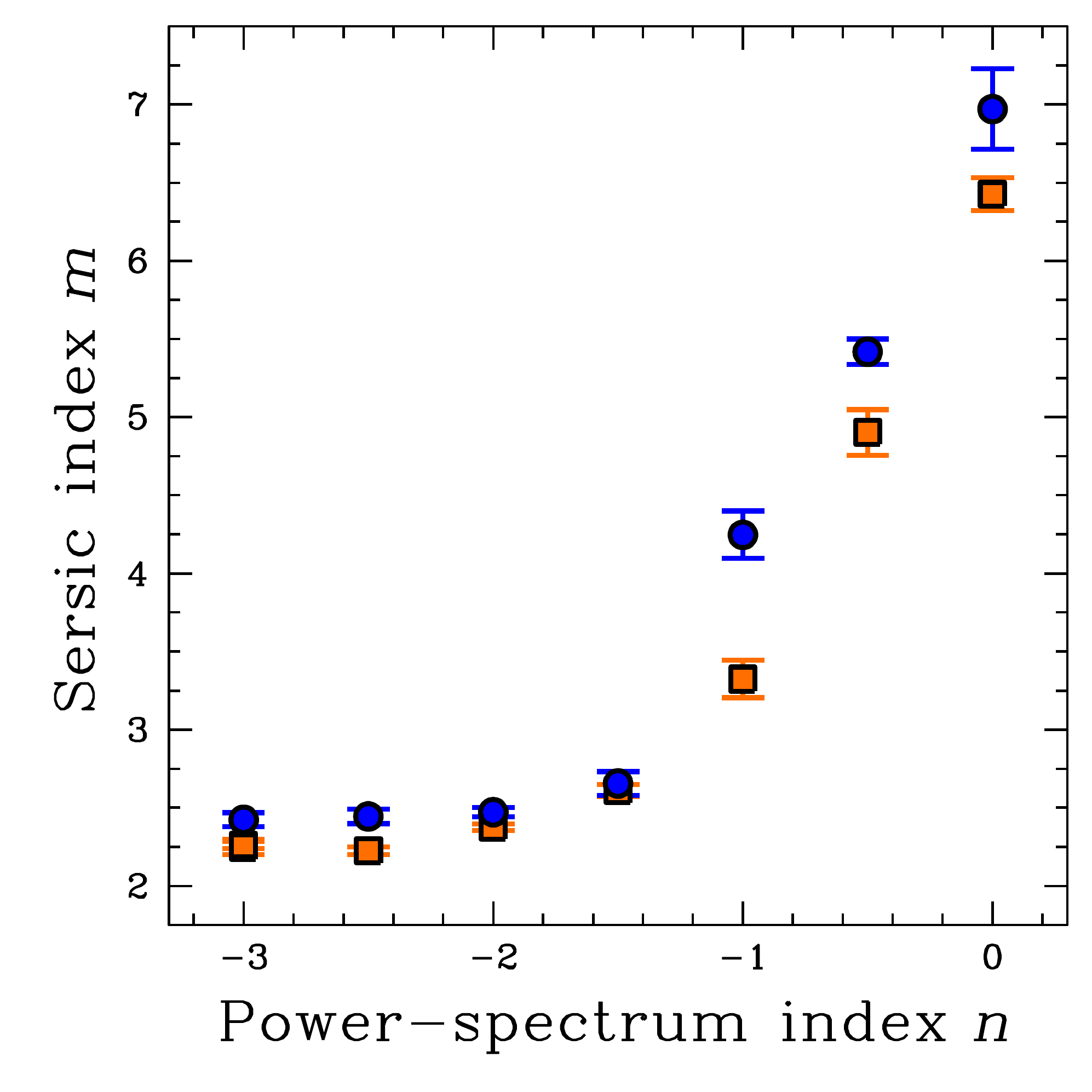} }
 \caption{Final best-fitting Einasto index $\alpha$ (left-hand panel)
   and deprojected {\Sersic} index $m$ (right-hand panel) as functions
   of the fluctuation power-spectrum index $n$ of the initial
   conditions of the $N$-body simulations. Pairs of simulations with
   the same value of $n$ differ only in the specific realization of
   the GRF (see Table~\ref{tab:par}). Error bars indicate 1-$\sigma$
   uncertainties.}
   \label{fig:ind}
\end{center}
\end{figure}

In \cite[Nipoti (2015, hereafter N15)]{Nip15} we have presented
idealized numerical experiments aimed at isolating the effect of the
fluctuation power spectrum shape on the density profile of a system
formed via cold dissipationless collapse. We refer the reader to N15
for details of the $N$-body simulations, in which the initial
conditions are realized by perturbing a smooth triaxial density
distribution with fluctuations generated from a GRF with power
spectrum $P(k)\propto k^n$, where $k$ is the wave-number.  Here we
present an additional set of simulations which, for given
power-spectrum index $n$, have all the same parameters as the
corresponding simulations of N15, but a different realization of the
GRF.  As done in N15 we fit the final angle-averaged density
distributions of the new realizations of the cold collapses with
deprojected {\Sersic} profiles (using the analytic approximation of
\cite[Lima Neto et al. 1999]{Lim99}). In addition we fitted the final
angle-averaged density distributions of all simulations (including
those of N15) with Einasto profiles, using as an estimate of the
half-mass radius $\rhalf$ the analytic approximation given by
\cite{Ret12}. In both cases we have only one free parameter (either
$m$ or $\alpha$) and we consider the radial range $0.04\leq
r/\rhalf\leq 10$.  Two examples of these fits are shown in
Fig.~\ref{fig:den}.  The best-fitting values of $\alpha$ and $m$ (and
1-$\sigma$ uncertainties) are reported in Table~\ref{tab:par} and
plotted in Fig.~\ref{fig:ind} as functions of the spectral index
$n$. We note that the deprojected {\Sersic} profile, though similar to
an Einasto profile, is not exactly an Einasto law, and that the
best-fitting $m$ of a deprojected {\Sersic} is not $1/\alpha$, where
$\alpha$ is the best-fitting Einasto index (\cite[Dhar \& Williams
  2010]{Dha10}).  The results of the simulations indicate that
$\alpha$ decreases (and $m$ increases) for increasing $n$: in other
words, high-$\alpha$ (low-$m$) systems form in smooth regions via few
mergers while low-$\alpha$ (high-$m$) systems form in clumpy regions
via several mergers, in agreement with the model of \cite{Cen14}.
Comparing the best-fitting index for two simulations with the same
value of $n$ it is apparent that different realizations of the same
GRF can lead to significant differences in the final values of
$\alpha$ and $m$ (see Fig.~\ref{fig:ind} and Table~\ref{tab:par}),
consistent with the relatively large scatter in the values of $\alpha$
measured, at fixed halo mass, in cosmological $N$-body simulations
(\cite[Dutton \& Macci{\`o} 2014]{Dut14}).

\begin{table}
  \begin{center}
\caption{Properties of the simulations. $n$: power-law index of the
  initial fluctuation power-spectrum. $\alpha$ and $m$: best-fitting
  Einasto and {\Sersic} indices of the final density profile ($\pm$
  1-$\sigma$). For each value of $n$ we have two simulations with
  different realizations of the GRF: (i) N15 (squares in
  Fig.~\ref{fig:ind}), (ii) this work (circles in
  Fig.~\ref{fig:ind}).\label{tab:par}}
\begin{tabular}{|r|r|r|r|r|}
\hline
$n$ & {$\alpha$ (i)~~~~~} &  {$\alpha$ (ii)~~~~~} & {$m$ (i)~~~~~} & {$m$ (ii)~~~~~}  \\
\hline
  -3  & $0.322\pm0.003$ & $0.292\pm0.007$ & $2.26\pm0.02$ & $2.42\pm0.05$ \\
  -2.5& $0.328\pm0.003$ & $0.295\pm0.006$ & $2.23\pm0.02$ & $2.45\pm0.05$ \\
  -2  & $0.302\pm0.002$ & $0.287\pm0.004$ & $2.38\pm0.02$ & $2.47\pm0.03$ \\
  -1.5& $0.275\pm0.005$ & $0.262\pm0.010$ & $2.61\pm0.04$ & $2.66\pm0.08$ \\
  -1  & $0.206\pm0.009$ & $0.161\pm0.006$ & $3.32\pm0.12$ & $4.25\pm0.15$ \\
  -0.5& $0.140\pm0.004$ & $0.127\pm0.002$ & $4.90\pm0.15$ & $5.42\pm0.08$ \\
     0& $0.106\pm0.002$ & $0.098\pm0.003$ & $6.43\pm0.11$ & $6.97\pm0.25$ \\
\hline
\end{tabular}
\end{center}
\end{table}

\end{document}